\pdfoutput=1

\documentclass[twocolumn,english,superscriptaddress]{revtex4}
\usepackage{amssymb,amsmath,graphicx}
\usepackage{color}
\usepackage[T1]{fontenc}
\usepackage{times}
\usepackage{babel}
\begin{document}



\title{
  Conductivity scaling and the effects of symmetry-breaking terms
  in bilayer graphene Hamiltonian 
}

\author{Dominik Suszalski}
\affiliation{Marian Smoluchowski Institute of Physics,
  Jagiellonian University, \L{}ojasiewicza 11, PL--30348 Krak\'{o}w, Poland}

\author{Grzegorz Rut}
\affiliation{Marian Smoluchowski Institute of Physics,
  Jagiellonian University, \L{}ojasiewicza 11, PL--30348 Krak\'{o}w, Poland}
\affiliation{Crif sp.\ z~o.o.,
  Lubla\'{n}ska 38, PL--31476 Krak\'{o}w, Poland}  

\author{Adam Rycerz}
\affiliation{Marian Smoluchowski Institute of Physics,
  Jagiellonian University, \L{}ojasiewicza 11, PL--30348 Krak\'{o}w, Poland}

\date{December 6, 2019}

\begin{abstract}
  We study the ballistic conductivity of bilayer graphene in the presence of 
  symmetry-breaking terms in effective Hamiltonian for low-energy excitations, 
  such as the trigonal-warping term ($\gamma_3$), the electron-hole symmetry 
  breaking interlayer hopping ($\gamma_4$), and the staggered potential
  ($\delta_{AB}$).
  Earlier, it was shown that for $\gamma_3\neq{}0$, in the absence of remaining
  symmetry-breaking terms (i.e., $\gamma_4=\delta_{AB}=0$), the conductivity
  ($\sigma$) approaches the value of $3\sigma_0$ for the system size
  $L\rightarrow{}\infty$ (with $\sigma_0=8e^2/(\pi{}h)$
  being the result in the absence of trigonal warping, $\gamma_3=0$).
  We demonstrate that $\gamma_4\neq{}0$ leads to the divergent
  conductivity ($\sigma\rightarrow{}\infty$) if $\gamma_3\neq{}0$, or to
  the vanishing conductivity ($\sigma\rightarrow{}0$) if $\gamma_3=0$.
  For realistic values of the tight-binding model parameters,
  $\gamma_3=0.3\,$eV, $\gamma_4=0.15\,$eV (and $\delta_{AB}=0$),
  the conductivity values are in the range of $\sigma/\sigma_0\approx{}4-5$ for
  $100\,$nm$\ <L<1\,\mu$m, in agreement with existing experimental results.
  The staggered potential ($\delta_{AB}\neq{}0$) suppresses zero-temperature
  transport, leading to $\sigma\rightarrow{}0$ for $L\rightarrow{}\infty$.
  Although $\sigma=\sigma(L)$ is no longer universal, the Fano factor
  approaches the pseudodiffusive value ($F\rightarrow{}1/3$ for
  $L\rightarrow{}\infty$) in any case with non-vanishing $\sigma$ (otherwise,
  $F\rightarrow{}1$) signaling the transport is ruled by evanescent waves.
  Temperature effects are briefly discussed
  in terms of a~phenomenological model for staggered potential
  $\delta_{AB}=\delta_{AB}(T)$ showing that, for
  $0<T\leqslant{}T_c\approx{}12\,$K and $\delta_{AB}(0)=1.5\,$meV, 
  $\sigma(L)$ is noticeably affected by $T$ for $L\gtrsim{}100\,$nm. 
\end{abstract}

\maketitle

\section{Introduction}
\label{intro}
Universal conductivity of monolayer graphene, $\sigma_{\rm MLG}=4e^2/(\pi{}h)$
(with the elementary charge $e$ and the Planck constant $h$), accompanied
by the pseudodiffusive shot noise (quantified by the Fano factor $F=1/3$), 
is one of the mosts recognizable landmarks of the Dirac's nature of
electrons dwelled in this material \cite{Kat06a,Two06,Mia07,Dic08,Dan08}.
These unique characteristics are link to a~dominant role of transport via
evanescent waves in graphene near the charge-neutrality point \cite{Kat12}.
What is more, the effective Hamiltonian for low-energy excitations, 
\begin{equation}
\label{hammlg}
  H_{\rm MLG}= v_F(p_x\sigma_x+p_y\sigma_y), 
\end{equation}
where $v_F=\sqrt{3}\,t_0{}a/(2\hbar)\approx{}10^6\,$m/s is the
energy-independent Fermi velocity (with $t_0\approx{}3\,$eV the
nearest-neighbor hopping integral and $a=0.246\,$nm the lattice spacing),
$p_j=-i\hbar{}\partial_j$ are in-plane momentum operators and $\sigma_j$ are
the Pauli matrices acting on sublattice degree of freedom (with $j=x,y$), 
possesses several symmetries which are crucial for the simplicity of transport
properties. (We further notice that the absence of valley-coupling factors
is supposed throughout the paper, and the discussion is limited to $K$ valley.)
These include: the {\em rotational invariance} (RI), the {\em electron-hole
symmetry} (EHS), and the {\em sublattice equivalence} (SE), which is embedded
in the so-called symplectic symmetry (or time-reversal symmetry in a~single
valley) \cite{Ben08,Wur09}. 
%

\begin{figure}
\centerline{
  \includegraphics[width=1.0\linewidth]{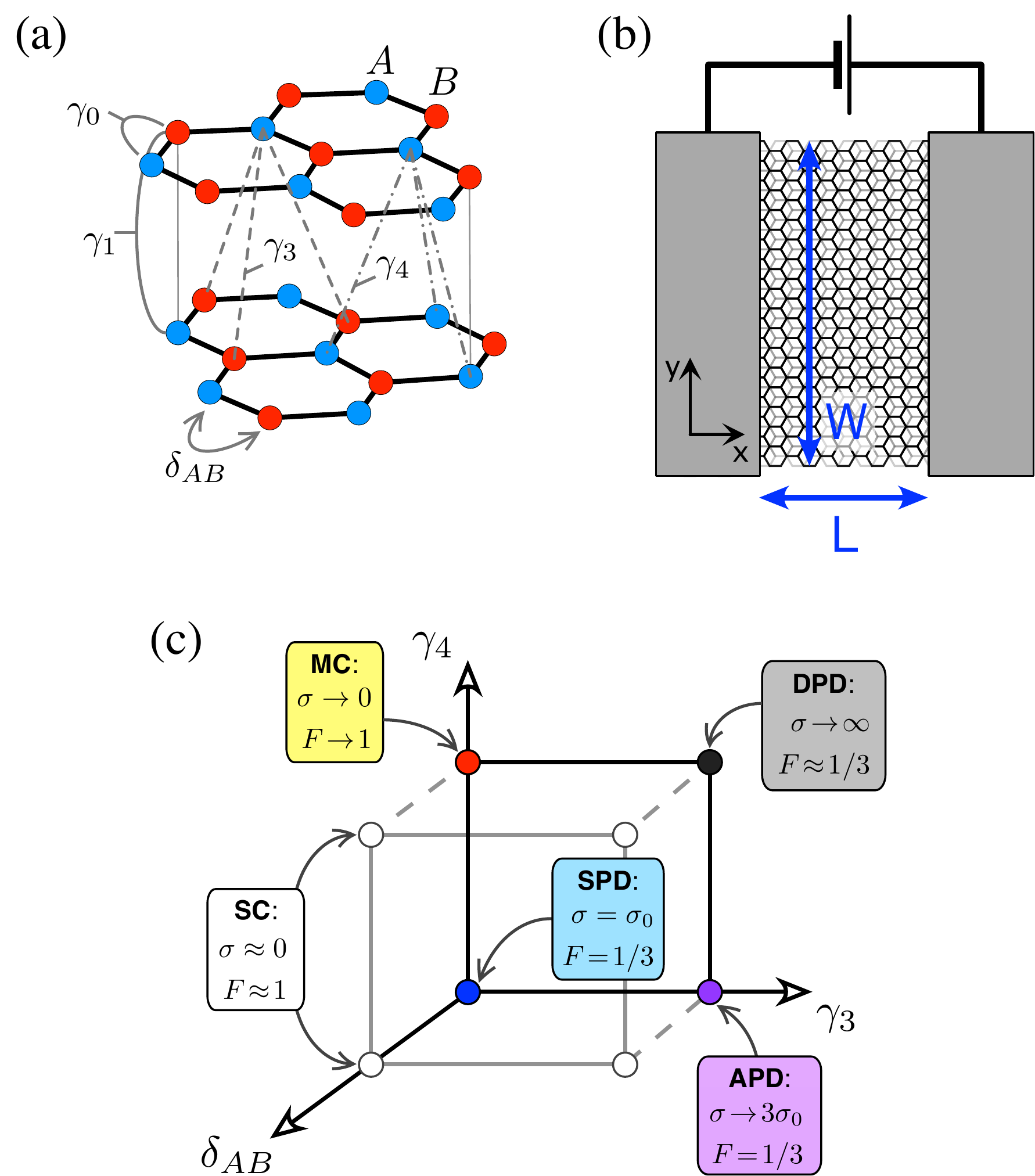}
}
\caption{ \label{outline}
(a) Tight-binding parameters for Bernal-stacked bilayer graphene,
(b) schematics of the system studied in the paper, and
(c) outline of our results for the conductivity $\sigma$ and the Fano
factor $F$. 
The limits of $\sigma$ and $F$ in panel (c) correspond
to $L\rightarrow{}\infty$ at a~fixed $W/L\gg{}1$, indicating
the following transport regimes: the standard pseudodiffusive (SPD),
the asymptotic pseudodiffusive (APD), the divergent pseudodiffusive (DPD),
the marginally conducting (MC), and the semiconducting (SC). 
Approximate equalities are used in the cases when the limiting values
are closely approached in the {\em mesoscopic} range of
$100\,$nm$\ \leqslant{}L\leqslant{}1\,\mu$m. 
}
\end{figure}

In bilayer graphene (BLG) the situation is more complex due to the couplings
between the layers \cite{Kat06b,Sny07,Cse07,Mog09,Rut14b}. 
Historically, the effective Hamiltonians for BLG were constructed by
taking only the leading tight-binding parameters 
of the Slonczewski-Weiss-McClure model \cite{Mcc57,Slo58} which are indicated
in Fig.\ \ref{outline}(a).

Even in the simplest possible approach \cite{Kat06b,Sny07}, including the
nearest-neighbor interlayer
hopping $\gamma_0$ (being numerically different then $t_0$) and the direct
interlayer hopping $\gamma_1$, SE is already eliminated due to inequivalence
of sites connected by $\gamma_1$ (dimer sites) and the remaining ones
(non-dimer sites), giving an opportunity to open the band gap by perpendicular
electric field introducing the layer inequivalence \cite{Mac13}.
(The second-neighbor interlayer hopping, formally breaking EHS, is
usually omitted as---in the low-energy limit---it only shifts the
charge-neutrality point by a~constant value; see Ref.\ \cite{Das11}.) 
Quite surprisingly, the approach of Refs.\ \cite{Kat06b,Sny07} leads to the
conductivity $\sigma_0=2\sigma_{\rm MLG}=8e^2/(\pi{}h)$ and $F=1/3$ (in the absence
of a~gap), as one could expect for two {\em decoupled} layers.
The results are also size-independent,
provided that $W\gg{}L\gg{}l_\perp$, with the sample width $W$, the length $L$
marked in Fig.\ \ref{outline}(b), and
$l_\perp=\sqrt{3}\,a\gamma_0/(2\gamma_1)\approx{}1.77\,$nm
being a~new length scale due the coupling between the layers.

Next, skew-interlayer hopping (or the {\it trigonal-warping} term) $\gamma_3$
\cite{gam2foo} breaks RI, leading to the appearance of three additional
Dirac cones at each valley \cite{Mac13}.
The effect of $\gamma_3$ on quantum transport is also significant
\cite{Cse07,Mog09,Rut14b}: Namely, the conductivity $\sigma(L)$ is no longer
universal but length-dependent, approaching the value of $3\sigma_0$ for
large $L$ \cite{anglefoo}.
In contrast, the Fano factor is unaffected (i.e., $F=1/3$), showing that
the pseudodiffusive nature of charge transport in BLG cannot be attributed 
any particular value of $\sigma$. 

In this paper, we complement the previous studies of ballistic charge
transport in BLG by examining numerically the effect of EHS-breaking
interlayer hopping $\gamma_4$ on the $\sigma$ (and $F$) dependence on $L$.
The results show that for $\gamma_4\neq{}0$, $\sigma(L)$ may either be
divergent (for $\gamma_3\neq{}0$) or vanishing (for $\gamma_3=0$) with
$L\rightarrow{}\infty$ (with $F\approx{}1/3$ in the first case or
$F\rightarrow{}1$ in the second case), as marked
schematically in Fig.\ \ref{outline}(c). 
These findings extend the collection of possible behaviors associated with
transport via evanescent waves in graphene-based systems. 
The role of {\it intristic} (i.e., unrelated to the external electric field
but rather interaction-induced) band gap reported by some experimental works
\cite{Bao12,Yan14,Gru15,Nam16} (and parametrized here by the staggered
potential $\delta_{AB}$) is also discussed. 

The remaining part of the paper is organized as follows. 
In Sec.~\ref{model} we present the model Hamiltonian and discuss how each
of the symmetry-breaking terms ($\gamma_3$, $\gamma_4$, or $\delta_{AB}$) 
affects the low-energy dispersion relation. 
Then, in Sec.~\ref{results} we demonstrate, by means of numerical mode-matching
for the Dirac equation, the behavior of $\sigma$ and $F$ with growing $L$
separately in the presence and in the absence of each symmetry breaking.
The concluding remarks are given in Sec.~\ref{conclu}. 

The numerical results presented in the main text are supplemented with
the explicit mode-matching
analysis for the special cases of $\gamma_3\neq{}0$, $\gamma_4=\delta_{AB}=0$
({\it Appendix~A\/}) and $\gamma_3=\delta_{AB}=0$, $\gamma_4\neq{}0$
({\it Appendix~B\/}).

\section{The model}
\label{model}

We start from the minimal version of the four-band Hamiltonian \cite{Mac13},
in which all the symmetry breakings mentioned in Sec.~\ref{intro} are
quantified by independent parameters  
\begin{equation}
\label{hamblg}
H_{\rm BLG} = \left(\begin{array}{cccc}
\delta_{AB}/2 & v_0\pi & \gamma_1 & -v_4\pi^\dagger \\
v_0\pi^\dagger & -\delta_{AB}/2 & -v_4\pi^\dagger & v_3\pi \\
\gamma_1 & -v_4\pi & -\delta_{AB}/2 & v_0\pi^\dagger \\
-v_4\pi & v_3\pi^\dagger & v_0\pi & \delta_{AB}/2  \\
\end{array}\right), 
\end{equation}
where $\pi=e^{-i\theta}(p_x+ip_y)$, $\pi^\dagger=e^{i\theta}(p_x-ip_y)$,
with the angle $\theta$ (between an armchair direction and the $x$-axis)
defining the crystallographic orientation of the sample, 
$v_0=\sqrt{3}a\gamma_0/\left(2\hbar\right)$,
$v_3=v_0\gamma_3/\gamma_0$, and $v_4=v_0\gamma_4/\gamma_0$.
In the forthcoming numerical discussion, we set $\theta=\pi/4$,
$\gamma_0=3.16\,$eV, and $\gamma_1=0.381\,$eV \cite{Kuz09};
for each of the remaining parameters 
the cases of zero- and nonzero-value are studied independently to demonstrate
the impact of a~particular symmetry breaking on ballistic transport.
Namely, we took $\gamma_3=0$ or $0.3\,$eV, $\gamma_4=0$ or $0.15\,$eV, and
$\delta_{AB}=0$ or $1.5\,$meV.

Our specific choice of the staggered potential $\delta_{\rm AB}$ in the
Hamiltonian $H_{\rm BLG}$ (\ref{hamblg}) follows from the demand that
it opens a~band gap without breaking EHS, which is solely controlled by
$\gamma_4$.
(In the parametrization of Ref.\ \cite{Mac13} the energy
difference between dimer- and non-dimer sites $V$ also breaks EHS, here
we set $V=0$). 
Physically, $\delta_{\rm AB}$ represents the irreducible part of a~gap
(i.e., that cannot be closed by external electric fields), and can be
attributed to charge or spin order which may appear in the BLG ground state
when electron-electron repulsive interactions are taken into account
\cite{Hir85,Thr14}. 

\begin{figure*}[!t]
  \includegraphics[width=0.9\linewidth]{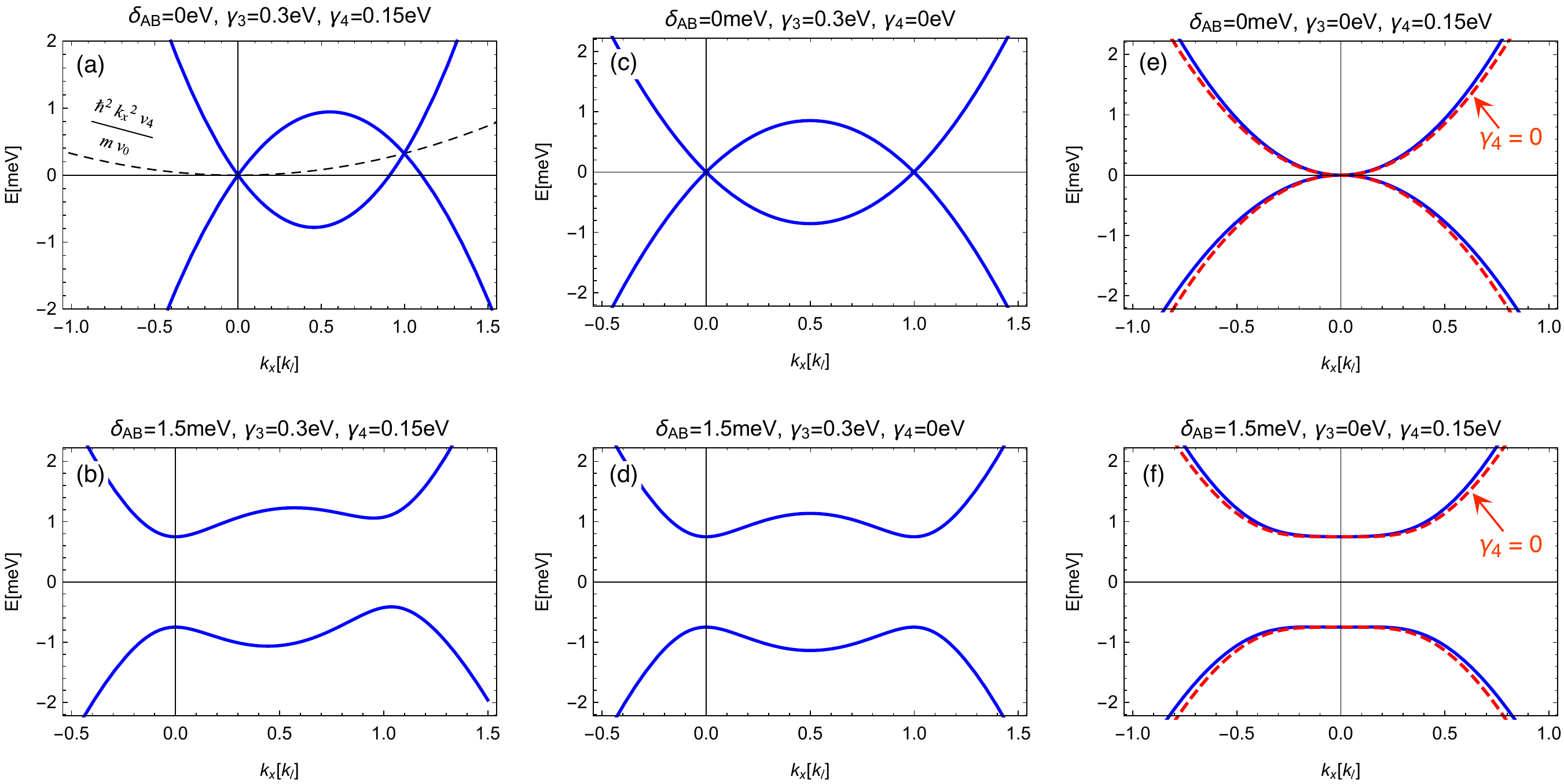}
\caption{ \label{dispersion}
  (a)--(f) Band energies for the Hamiltonian $H_{\rm BLG}$ given by Eq.\
  (\ref{hamblg}) in the main text with $\theta=0$. 
  Solid lines correspond to the parameters $(\delta_{AB},\gamma_3,\gamma_4)$
  specified in each panel.
  Black dashed line in panel (a) marks the parabolic correction due to
  the electron-hole symmetry breaking interlayer hopping $\gamma_4$
  (see the explicit formula with $m=\gamma_1/2v_0^2$).
  Red dashed lines in panels (e), (f) depict the reference band structure
  for $\gamma_3=\gamma_4=0$. The wavenumber $k_x=p_x/\hbar$ is specified in
  units of $k_l=\frac{2}{3}\sqrt{3}\,\gamma_1\gamma_3/(a\gamma_0^2)
  \approx{}0.05\,$nm$^{-1}$,
  being the $k_x$-position of a~secondary Dirac cone calculated for the
  parameters as listed in panel (c).
  Each panel displays the cross section taken at $k_y=0$. 
}
\end{figure*}

In Fig.\ \ref{dispersion} we present low-energy band structure following from
the Hamiltonian $H_{\rm BLG}$ (\ref{hamblg}) by displaying the cross sections,
for $p_y=\hbar{}k_y=0$, of dispersion relations for eight different combinations
of symmetry-breaking parameters $\gamma_3$, $\gamma_4$, and $\delta_{AB}$.
An apparent feature visible in Fig.\ \ref{dispersion}(a) is the energy shift
of a~secondary Dirac cone (same for all three secondary cones) due to EHS
breaking for $\gamma_4\neq{}0$
[see Fig.\ \ref{dispersion}(c) for a~comparison] making impossible
(for $\gamma_3\neq{}0$ and $\delta_{AB}=0$) to achieve the exact zero-doping
case, in which the transport is fully carried by evanescent waves. 
In contrast, for $\gamma_3=0$ [see Figs.\ \ref{dispersion}(e) and
\ref{dispersion}(f)], the effects of $\gamma_4$ are marginal (apart from
clear electron-hole asymmetries visible for $\gamma_4\neq{}0$), and the
zero-doping case can be achieved for both $\gamma_4=0$ or $\gamma_4\neq{}0$. 
For $\delta_{AB}\neq{}0$, we have an indirect band gap for $\gamma_4\neq{}0$
and $\gamma_4\neq{}0$ [see Fig.\ \ref{dispersion}(b)], or direct bandgaps
in the remanining cases, allowing one to obtain the zero doping by adjusting
the Fermi level to the gap. 

Consequences of these features for BLG transport properties
are discussed next.

\section{Results and discussion}
\label{results}

\subsection{Zero-temperature charge transport}
We employ the Landauer-B\"{u}ttiker expressions for zero-temperature
conductivity and the Fano factor in the linear-response regime
\cite{Naz09}, namely 
\begin{align}
  \sigma(T\!\rightarrow{}\!0) &= \frac{g_0L}{W}
  \mbox{Tr}\left({\bf t}{\bf t}^\dagger\right), \label{sigzerte} \\
  F &= \frac{\mbox{Tr}\left[\,{\bf t}{\bf t}^\dagger %
  \!\left({\bf 1}\!-\!{\bf t}{\bf t}^\dagger\right)\right]}%
  {\mbox{Tr}\left({\bf t}{\bf t}^\dagger\right)},
\end{align}
with the conductance quantum $g_0=4e^2/h$ accounting for spin and valley
degeneracies. 
In order to determine the transmission matrix at a~given Fermi energy,
${\bf t}={\bf t}(E)$, for a rectangular sample attached to the two
heavily-doped regions, we employ the computational scheme similar to
the presented in Ref.\ \cite{Rut14b}, with a~numerical stabilization
introduced in Ref.\ \cite{Sus18}.
In brief, at finite-precision arithmetics, the mode-matching equations may
become ill defined for sufficiently large $L$, as they contain both
exponentially growing and exponentially decaying coefficients.
This difficulty is overcome by dividing the sample area into $N_{\rm div}$
consecutive, equally-long parts, and matching the wave functions for all
(i.e., $N_{\rm div}+1$) interfaces. Typically, using the double-precision
arithmetic, we put $N_{\rm div}=\lfloor{}L/(40\,l_\perp)\rfloor+1$, 
with $\lfloor{}x\rfloor$ denoting the floor of $x$. 

\begin{figure*}[!t]
  \includegraphics[width=0.7\linewidth]{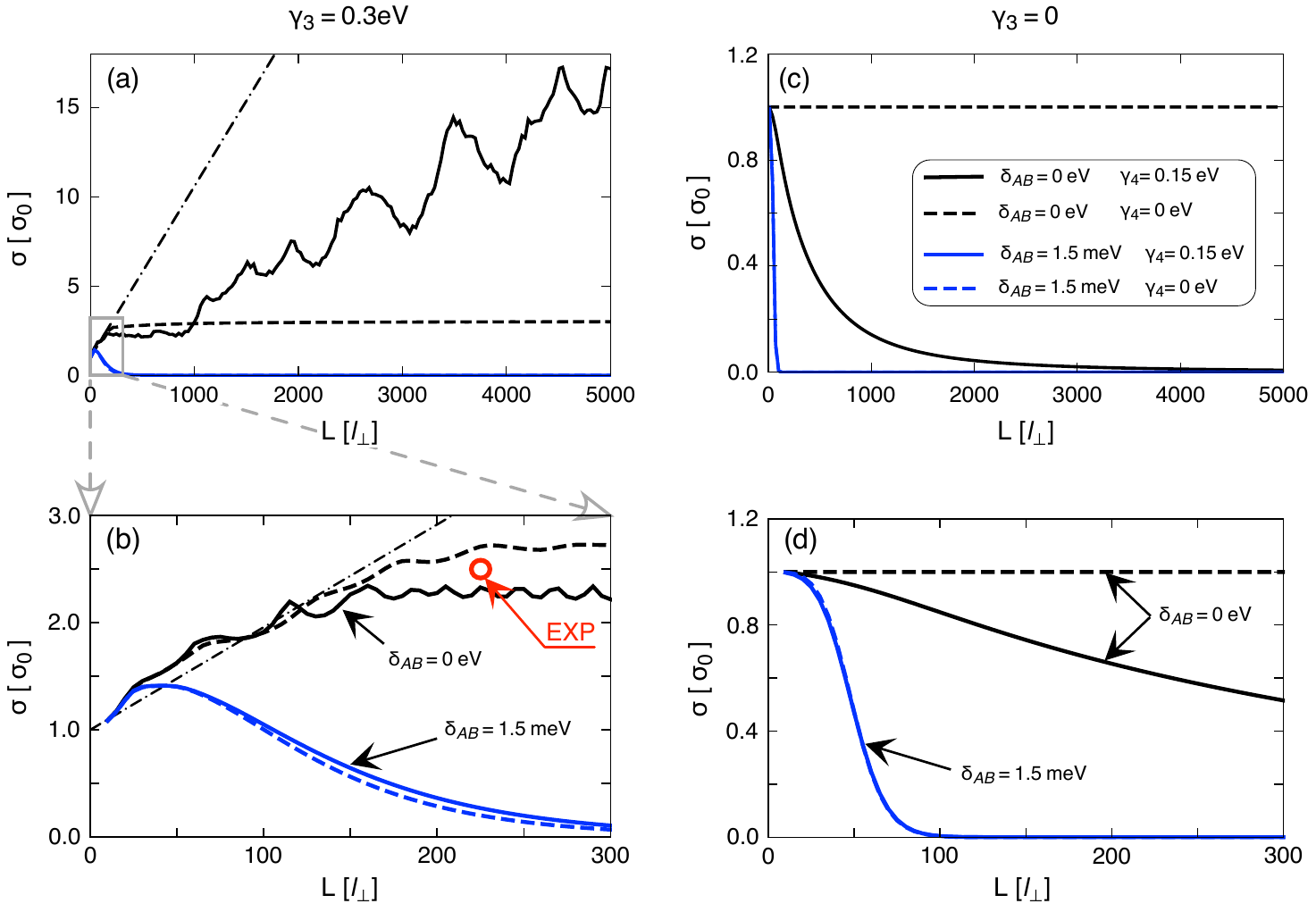}
\caption{ \label{conductivity}
  Conductivity (in the units of $\sigma_0=8e^2/(\pi{}h)$) as a~function of
  length $L$ (in the units of $l_\perp=\hbar{}v_0/\gamma_1=1.77\,$nm)
  at a~fixed $W/L=20$.
  The trigonal warping strength is $\gamma_3=0.3\,$eV (a,b) or
  $\gamma_3=0$ (c,d). Bottom panels are zoom-ins, for small $L$, of the data
  presented in top panels.
  Solid (or dashed) lines in all panels are for $\gamma_4=0.15\,$eV
  (or $\gamma_4=0$); the staggered potential is $\delta_{AB}=0$ (black lines),
  or $\delta_{AB}=1.5\,$meV (blue lines), as indicated with arrows in panel (b)
  and (d). Red circle in panel (b) marks the experimental results of
  Ref.\ \cite{May11}.
  Dash-dot lines in panels (a,b) depict the approximate upper bound given by
  Eq.\ (\ref{sigmabound}) in the main text. 
  (The line-color encoding shown in panel (c) is same for all panels.) 
}
\end{figure*}

\begin{figure*}[!t]
  \includegraphics[width=0.7\linewidth]{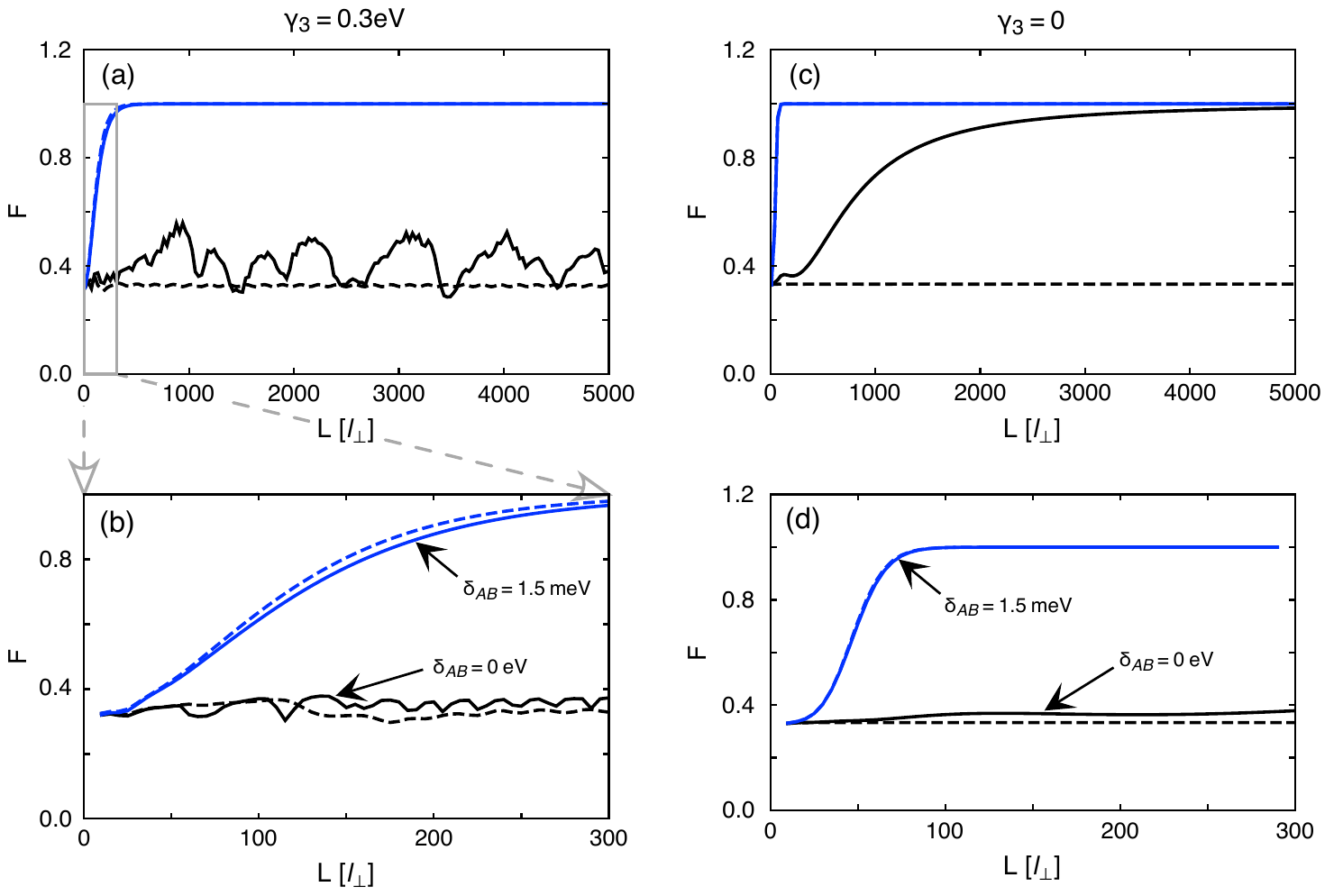}
\caption{ \label{fanofactor}
  (a)--(d) The Fano factor $F$ as a~function of $L$. The system parameters
  and the line-color encoding are same as in Fig.\ \ref{conductivity}. 
}
\end{figure*}

Our numerical results for $E=0$ are presented in Figs.\ \ref{conductivity}
and \ref{fanofactor}.
As the debate on the ground-state nature in BLG is currently ongoing
\cite{Thr14} and the existing experimental results are far from being
consistent \cite{Bao12,Yan14,Gru15,Nam16,May11,Kra18},
we examine eight possible scenarios by setting  different values of 
the parameters $(\delta_{AB},\gamma_3,\gamma_4)$ in the low-energy Hamiltonian
$H_{\rm BLG}$ (\ref{hamblg}),
corresponding to the dispersion relations presented in Sec.~\ref{model}. 

The behavior of transport properties is relatively simple for 
$\delta_{\rm AB}=1.5\,$meV (coinciding with the gap reported by Refs.\
\cite{Gru15,Nam16}), we observe a~fast decay of $\sigma(L)$ with growing $L$,
accompanied by $F\rightarrow{}1$
(see {\em blue lines} in Figs.\ \ref{conductivity} and \ref{fanofactor}),
indicating the insulating (or semiconducting) behavior. 
The remaining parameters ($\gamma_3$ and $\gamma_4$) are essentially
meaningless in such a~case; a~slightly elevated conductivity (namely, 
$\sigma>\sigma_0$) is visible for $\gamma_3=0.3\,$eV and $L<100\,l_\perp$,
due to the finite-size effects. 

In a~gapless case ($\delta_{\rm AB}=0$) we identify four apparently different
behaviors of $\sigma(L)$, depending on whether the remaining
parameters ($\gamma_3$ and $\gamma_4$) take zero- or non-zero values.

For $\delta_{\rm AB}=0$ and $\gamma_3=0.3\,$eV, the values of $\sigma(L)$ are
generically elevated (above $\sigma_0$) for any $L$
[{\em black lines} in Figs.\ \ref{conductivity}(a) and \ref{conductivity}(b)],
with the Fano factor $F\approx{}1/3$ 
[{\em black lines} in Figs.\ \ref{fanofactor}(a) and \ref{fanofactor}(b)]. 
Starting from $L\approx{}2000\,l_\perp$, one can further distinguish
the behaviors for $\gamma_4=0.15\,$eV [{\em black solid} line
in Fig.\ \ref{conductivity}(a)], for which $\sigma(L)$ grows approximately
linearly with $L$, and for $\gamma_4=0$ [{\em black dashed} line],
for which $\sigma(L)$ approaches the value of $3\sigma_0$.

For $\gamma_4=0.15\,$eV, the energy shift of three secondary Dirac cones
[see Fig.\ \ref{dispersion}(a)] is equal to
\begin{equation}
\label{delel}
  \Delta{}E_l=\frac{\hbar^2k_l^2v_4}{mv_0}=
  2\frac{\gamma_1\gamma_3^2\gamma_4}{\gamma_0^3}\approx{}0.33\,\text{meV}, 
\end{equation}
where $m=\gamma_1/2v_0^2$ and $k_l=\gamma_1v_3/\hbar{}v_0^2$, leading to
a~nonzero number of propagating modes ({\em open channels}) at zero energy,
which can be approximated as \cite{Sus18} 
\begin{equation}
  N_{\rm open}(E\!=\!0)\approx 0.68\,\frac{\Delta{}E_l{}\,W}{\hbar{}v_3}.  
\end{equation}
Subsequently, the excess conductivity from secondary Dirac cones 
can roughly be bounded as 
\begin{equation}
\label{sigmabound}
\sigma(L) -\sigma_0 \lesssim \frac{g_0{}N_{\rm open}L}{W}=
9.6\cdot{}10^{-3}\,\sigma_0\frac{L}{l_\perp}, 
\end{equation}
with the rightmost equality corresponding to $W/L=20$ and
$\sigma_0$ on the left-hand side representing a~contribution from
evanescent waves in the primary Dirac cone;  see {\em dash-dot} lines
in Figs.\ \ref{conductivity}(a) and \ref{conductivity}(b). 
The transmission reduction for propagating modes, approximately by a~factor
of $2$, can be attributed to the additional backscattering appearing in the
double-barrier geometry, which is usually much weaker for a~single barrier
\cite{Sus19}. 
A~secondary feature of $\sigma(L)$ is a~quasiperiodic oscillation due to the
Fabry-Perrot resonances appearing for $L=n\Delta{}L$, where
$n=1,2,\dots$, and (up to the order of magnitude) 
$\Delta{}L\sim\pi\hbar{}v_3{}/\Delta{}E_l \approx 340\,l_\perp$ \cite{Sus18}.   

None of these effects is present for $\gamma_4=0$, for which the conductivity
follows the scenario earlier described in Refs.\ \cite{Mog09,Rut14b}.
(In {\it Appendix~A\/}, we present the analytical derivation explaining why
$\sigma(L)\rightarrow{}3\sigma_0$ for $L\rightarrow{}\infty$ and arbitarily
small $\gamma_3\ne{}0$.) 
We further notice that the available experimental value Ref.\ \cite{May11},
reporting $\sigma\approx{}2.5\sigma_0$ for $L\approx{}400\,$nm$\ =226\,l_\perp$ 
[{\em red circle} in Fig.\ \ref{conductivity}(b)], is equally close to both
the results for $\gamma_4=0$ and $0.15\,$eV, and the determination of
$\gamma_4$ via conductivity measurements requires a~sample length exceeding
$L\gtrsim{}2\,\mu$m. 

For $\delta_{\rm AB}=\gamma_3=0$, the conductivity behavior with growing $L$ 
is a~bit more peculiar.

If $\gamma_4=0$, we simply have $\sigma(L)=\sigma_0$ and $F=1/3$
for any $L\gg{}l_\perp$ [see {\em black dashed} lines in Figs.\
\ref{conductivity}(c,d) and \ref{fanofactor}(c,d)], 
reproducing the analytical results of Refs.\ \cite{Kat06b,Sny07}. 

In contrast, if $\gamma_4=0.15\,$eV [{\em black solid} lines] we observe
a~slow power-lay decay of $\sigma(L)$ with growing $L$, which can be
approximated as $\sigma(L)\propto{}L^{-2.0}$ for $L\gtrsim{}1000\,l_\perp$,
accompanied by $F\rightarrow{}1$.
Notice that the Fano factor is $F\approx{}1/3$ in the range of
$L\leqslant{}300\,l_\perp$ shown in Fig.\ \ref{fanofactor}(d);
the convergence to $1$ becomes visible for $L\gtrsim{}1000\,l_\perp$,
see Fig.\ \ref{fanofactor}(c). 
Unlike for nonrelativistic electrons \cite{nonrelfoo} we still obtain
a~finite $\sigma(L)$ in the limit of infinite doping in the leads
at fixed $W$ and $L$, signaling the relativistic nature of charge carriers. 
The vanishing conductivity for $L\rightarrow{}\infty$ at a~fixed
$W/L$, in the absence of a~gap, clearly represents a~remarkable feature
of the results, providing an opportunity to verify the $\gamma_3=0$ model
as put forward in Ref.\ \cite{Yan14} within ballistic transport experiments. 
A~further reasoning that such a~behavior appears generically
for $\gamma_3=0$ and $\gamma_4\neq{}0$, is given in {\it Appendix~B\/}.

\subsection{Finite-temperature effects}

\begin{figure*}[!t]
  \includegraphics[width=0.7\linewidth]{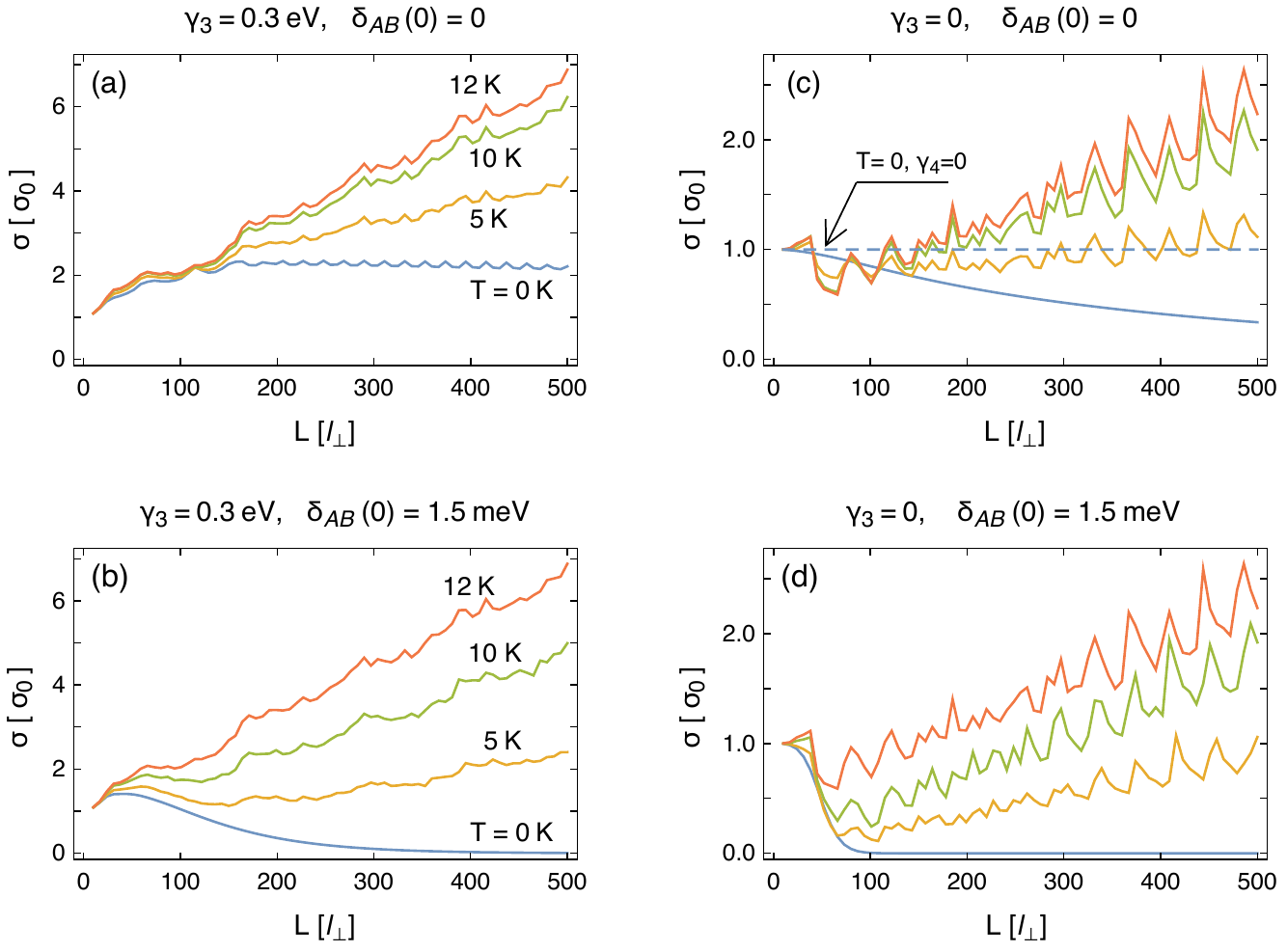}
\caption{ \label{condtemp}
  (a)--(d) Conductivity as a~function of $L$ for temperatures
  $T=0\,K$, $5\,K$, $10\,K$, and $T=T_C=12\,K$ [see Eq.\ (\ref{deltemp}) in
  the main text] indicated in panels (a,b) and same for all panels.
  The parameters $\gamma_3$ and $\delta_{AB}(0)$ are specified for each panel;
  the parameter $\gamma_4=0.15\,$eV for all lines, with the exception of
  dashed line for $T=0$ in panel (c), for which $\gamma_4=0$
  (as marked with an~arrow). 
}
\end{figure*}

For $T>0$ and in the linear-response regime the electronic noise is
dominated by the Nyquist-Johnson term of $S(0)\approx{}4k_BT{}\sigma{}W/L$
\cite{Naz09}, and the Fano factor becomes irrelevant. Therefore,
we limit our discussion to the temperature-dependent conductivity,
which is given by 
\begin{equation}
\label{sigfinte}
  \sigma(T\!>\!0) =
  \frac{g_0L}{W}\int{}dE\ \mbox{Tr}\left({\bf t}{\bf t}^\dagger\right)
  \left(-\frac{\partial{}f_{\rm FD}}{\partial{}E}\right), 
\end{equation}
where $f_{\rm FD}(\mu,T,E)=\left[\,\exp((E-\mu)/k_BT)+1\,\right]^{-1}$ is
the Fermi-Dirac distribution function for a~given chemical potential $\mu$,
and the remaining symbols are same as in Eq.\ (\ref{sigzerte}).

Numerical integration in Eq.\ (\ref{sigfinte}) is performed, for $\mu=0$,
by taking the energy range of $-E_{\rm M}\leqslant{}E\leqslant{}E_{\rm M}$,
with a~cut-off energy $E_{\rm M}=0.05\,$eV
(i.e., $E_M >48\,k_BT$ for $T\leqslant{}12\,$K) 
being sufficiently high to reach a~convergence up to the machine round-off
errors. 
Additionally, when calculating the transmission matrix ${\bf t}(E)$,
we parametrize the staggered potential in the effective Hamiltonian
$H_{\rm BLG}$~(\ref{hamblg}) as follows
\begin{multline}
\label{deltemp}
  \delta_{AB}(T)= \delta_{AB}(0) \\
  \times
  \begin{cases}
    \tanh\left(1.74\,\sqrt{\frac{T_C}{T}-1}\,\right)
    & \text{if }\ T\leqslant{}T_C \\
    0 & \text{if }\ T>T_C
  \end{cases}, 
\end{multline}
with $T_C=12\,$K and $\delta_{AB}(0)=1.5\,$meV reproducing the temperature
dependence of a~gap reported in Refs.\ \cite{Gru15,Nam16}.
(The gapless case $\delta_{AB}(0)=0$ is considered separately.) 

Our numerical results,  for $T=0$ and the selected temperatures 
$0<T\leqslant{}T_C$, are presented in Fig.\ \ref{condtemp}.
Similarly as in the previous subsection, the datasets for $\gamma_3=0.3\,$eV
[see Figs.\ \ref{condtemp}(a) and \ref{condtemp}(b)] and for $\gamma_3=0$
[see Figs.\ \ref{condtemp}(c) and \ref{condtemp}(d)] are displayed 
separately. This time, we limit the presentation to the $\gamma_4=0.15\,$eV
case for clarity ({\em solid lines} in all panels), as the curves for
$\gamma_4=0$ closely follows their $\gamma_4=0.15\,$eV counterparts,
with the exception for $\gamma_3=\delta_{AB}(0)=0$ and $T=0$ [see {\em dashed
line} in Fig.\ \ref{condtemp}(c)], when the conductivity suppression 
in the presence of EHS symmetry breaking ($\gamma_4\neq{}0$) is clearly
visible.

An apparent feature of the results presented in Fig.\ \ref{condtemp}(a) 
is that the curves for different temperatures closely follow each other
up to $L\lesssim{}120\,l_\perp$, above which $\sigma(L)$ grows noticeably
faster with $L$ for higher $T$. [Notice that the $T=0$ curve also shows
approximately linear growths with $L$, which manifests itself for
$L\gtrsim{}10^3\,l_\perp$; see Fig.\ \ref{conductivity}(a).]
The position of a~coalescence point, $L\approx{}120\,l_\perp$, can be attributed
to fact that above such a~length the quantum-size effects are less significant,
allowing the finite-temperature effects to dominate transport properties. 
This can be rationalized taking into account the time-energy uncertainty
relation limiting the energy resolution 
\begin{equation}
  \delta{}E\geqslant{}\frac{\hbar}{2\tau_{\rm flight}}=\frac{\hbar{}v_3}{2L}, 
\end{equation}
with $\tau_{\rm flight}\approx{}L/v_3$ being the ballistic time of flight
\cite{Sus18}, together with the fact that the energy of thermal excitations is
$k_BT\gtrsim{}\Delta{}E_l$ for $T\gtrsim{}4\,$K, with $\Delta{}E_l$ given by
Eq.\ (\ref{delel}). 
Subsequently, one can expect that the propagating modes in secondary Dirac
cones are employed (by thermal excitations) provided that
$\delta{}E\ll\Delta{}E_l\lesssim{}k_BT$, leading to
\begin{equation}
\label{lenboundv3}
  \frac{L}{l_\perp}\gg{}\frac{\gamma_0^2}{4\gamma_3\gamma_4}\approx{}55. 
\end{equation}

For $\gamma_3=0$ the above reasoning no longer applies, however,
a~relatively flat $\sigma$ dependence on $L$ for $T>0$
[see Fig.\ \ref{condtemp}(c)] coincides with the divergent lower bound
for $L$ in Eq.\ (\ref{lenboundv3}).
In such a~case, one should rather estimate the time of flight (up to the order
of magnitude) as $\tau_{\rm flight}\sim{}L/v_0$. In turn, the condition
$k_BT\gg\delta{}E$ allowing the conductivity enhancement by thermal excitations
is equivalent to
\begin{equation}
\label{lenboundv0}
  \frac{L}{l_\perp} \gg{} \frac{\gamma_1}{2k_BT}\approx{}
  \frac{2200\,\text{K}}{T}, 
\end{equation}
giving, for instance, $L\gg{}440\,l_\perp$ for $T=5\,$K. 
The lower bound for $L$ in Eq.\ (\ref{lenboundv0}) allows to understand
why temperature effects on $\sigma(L)$ are noticeably weakened for
$\gamma_3=0$, comparing to the $\gamma\neq{}0$ case. 

In the presence of a~staggered potential, $\delta_{\rm AB}(0)=1.5\,$meV,
the primary temperature effects on $\sigma(L)$ visible in
Figs.\ \ref{condtemp}(b) and \ref{condtemp}(d) can be attributed to the gap
closing for $T$ approaching $T_C$; see Eq.\ (\ref{deltemp}).
The characteristic system length, above which the value of $\delta_{AB}$
becomes significant, derived from the condition for the energy uncertainty
$\delta{}E\ll{}\delta_{AB}(0)$, reads 
\begin{equation}
\label{lenbounddel0}
  \frac{L}{l_\perp} \gg
    \begin{cases}
    \frac{1}{2}\gamma_1\gamma_3/\gamma_0\delta_{AB}(0) \approx{}12.1
    & \text{if }\ \gamma_3\neq{}0 \\
    \frac{1}{2}\gamma_1/\delta_{AB}(0) \approx{}127 & \text{if }\ \gamma_3=0 
  \end{cases}, 
\end{equation}
where we have estimated $\tau_{\rm flight}\approx{}L/v_3$ (if $\gamma_3\neq{}0$)
or $\tau_{\rm flight}\sim{}L/v_0$ (if $\gamma_3=0$).
This time, our numerical results show that the temperature
effects are visible for significantly shorter systems 
in the $\gamma_3=0.3\,$eV case, comparing to the $\gamma_3=0$ case,
in a~qualitative agreement with the estimation given in
Eq.\ (\ref{lenbounddel0}). 

\section{Concluding remarks}
\label{conclu}

We have investigated, by calculating ballistic transport characteristics
within the Landauer-B\"{u}ttiker formalism, the role of symmetry-breaking
terms in the effective Hamiltonian for bilayer graphene.
Three of such terms, the trigonal warping ($\gamma_3$), the electron-hole
symmetry breaking interlayer hopping ($\gamma_4$), and the staggered
potential ($\delta_{AB}$) quantifying a~{\em spontaneous} band gap, 
are independently switched on and off, resulting in different behaviors
of the conductivity ($\sigma$) and the Fano factor ($F$) with the increasing
system length ($L$) at a~fixed width-to-length ratio ($W/L$). 

In the absence of a~gap ($\delta_{AB}=0$) one can identify three different
quantum-transport regimes characterized by the pseudodiffusive shot-noise
power, $F=1/3$:
(i) the {\em standard pseudodiffusive} regime, characterized by
$\sigma(L)=\sigma_0$ (with $\sigma_0=8e^2/(\pi{}h)$ being a~double
conductivity of a~monolayer) and occurring for $\gamma_3=\gamma_4=0$,
(ii) the {\em asymptotic pseudodiffusive} regime, with
$\sigma(L)\rightarrow{}3\sigma_0$ for $L\rightarrow{}\infty$, occuring for
$\gamma_3\neq{}0$ and $\gamma_4=0$, and
(iii) the {\em divergent pseudodiffusive} regime, with
$\sigma(L)\rightarrow{}\infty$ for $L\rightarrow{}\infty$, occuring for
$\gamma_3\neq{}0$ and $\gamma_4\neq{}0$.
Additionally, for $\gamma_3=0$ and $\gamma_4\neq{}0$, the system can be
regarded as a~marginal conductor, with $\sigma(L)\rightarrow{}0$
(showing a~power-law decay)
and $F\rightarrow{}1$ for $L\rightarrow{}\infty$.


In the presence of a~staggered potential at $T=0$ ($\delta_{AB}(0)>0$),
a~semiconducting
behavior is observed regardless the remaining parameters ($\gamma_3$ and
$\gamma_4$); i.e., $\sigma(L)\rightarrow{}0$ (showing the exponential decay)
and $F\rightarrow{}1$ for $L\rightarrow{}\infty$.
For $T>0$, a~zero-gap behavior
is gradually restored (for any combination of $\gamma_3$ and $\gamma_4$)
when the energy of thermal excitations 
$k_BT\gtrsim{}\delta_{AB}(0)$. 

We hope that our numerical results will help verifying the bilayer graphene
models proposed in the literature, as soon as ballistic samples of the length
$L\gtrsim{}1\,\mu$m become available. So far, conductivity measurements for
shorter samples \cite{May11} suggest that the models neglecting the trigonal
warping ($\gamma_3=0$) cannot correctly reproduce transport properties in
the mesoscopic range, but a~conclusive information concerning
the value of $\gamma_4$ is missing.

Apart from the material-science aspects outlined above, the asymptotic
conductivity behavior suggests that bilayer graphene represents a~model
case when discussing the generality of spontaneous symmetry breaking
in quantum systems \cite{Wez07,Wez06}.
When $\sigma$ is considered as an order parameter, our findings can be
summarized by putting forward the non-commuting order of limits,
as $L\rightarrow\infty$ and the relevant symmetry-breakings vanish; namely
\begin{align}
  \lim_{L\rightarrow{}\infty}\lim_{d\rightarrow{}\infty}\lim_{\delta_{AB}\rightarrow{}0}
  \sigma  &= 
  \lim_{L\rightarrow{}\infty}\lim_{\delta_{AB}\rightarrow{}0}\lim_{d\rightarrow{}\infty}
  \sigma = \sigma_0, \\
  \lim_{d\rightarrow{}\infty}\lim_{L\rightarrow{}\infty}\lim_{\delta_{AB}\rightarrow{}0}
  \sigma & = \infty, \\
  \lim_{d\rightarrow{}\infty}\lim_{\delta_{AB}\rightarrow{}0}\lim_{L\rightarrow{}\infty}
  \sigma &=
  \lim_{\delta_{AB}\rightarrow{}0}\left[\,\dots\right]\,
  \sigma = 0, \label{limdelab}
\end{align}
where we have introduced the distance between the layers $d$
(with $d\rightarrow\infty$ corresponding to simultaneous limits of 
$\gamma_3\rightarrow{}0$ and $\gamma_4\rightarrow{}0$ \cite{limdfoo}),
and the dots $[\,\dots]$ in Eq.\ (\ref{limdelab}) mark that the order of
the two remaining limits is arbitrary in such a~case.
From this perspective, it becomes clear that both the subblattice- and
the combined rotational-electron-hole symmetry breakings
may appear spontaneously, as consequences of the layer stacking in graphene
($d=\text{const}<\infty$). 

The peculiar cases of $\gamma_3\neq{}0$ or $\gamma_4\neq{}0$ in the
absence of other symmetry breakings (i.e., $\delta_{AB}=\gamma_4=0$ or
$\delta_{AB}=\gamma_3=0$) do not seem to have as clear physical 
interpretations. 
However, in heterostructures containing graphene, 
a~variety of spontaneuous symmetry breakings may appear due to
the couplings to surrounding layers, encouraging one to consider
also anomalous parameter configurations.


\section*{Acknowledgements}

The work was supported by the National Science Centre of Poland (NCN)
via Grant No.\ 2014/14/E/ST3/00256. Computations were partly performed
using the PL-Grid infrastructure.


\appendix

\section{
  Transmission through bilayer graphene in the presence of trigonal warping
  ({\boldmath$\gamma_3\neq{}0$}, {\boldmath$\gamma_4=0$}) 
}

Here, we present the analytical derivation of the {\em total transmision}
(i.e., transmission probability summed over normal modes), coinciding
with the Landauer-B\"{u}ttiker conductivity [see Eq.\ (\ref{sigzerte}) in the
main text] 
$\sigma(L)\rightarrow{}3\sigma_0$
in the limit of $L,W\rightarrow{}\infty$, at $W/L=\text{const}\gg{}1$. 
Some partial results were earlier reported in
Ref.\ \cite{Mog09}, but the full derivation, to our best knowledge, is
missing in the literature.

The disperion relation for the Hamiltonian given by Eq.\ (\ref{hamblg})
in the main text, with $\delta_{AB}=\gamma_4=0$, takes a~form
\begin{align}
  E^2 &= \frac{\gamma_1^2}{2} +\left( v_0^2 + \frac{v_3^2}{2}\right) p^2 \pm
  \sqrt{\Gamma}, \label{disp0a} \\
  \Gamma &= \frac{1}{4} \left( \gamma_1^2-v_3^2 p^2 \right)^2 +v_0^2 p^2
  \left(\gamma_1^2 + v_3^2 p^2\right) \nonumber \\
  &+ 2\gamma_1 v_3 v_0^2 p^3 \cos(3\varphi), \label{disp0b}
\end{align}
where $p=\sqrt{p_x^2+p_y^2}\,$ and we have set $\theta=0$ for simplicity
(later, we show that the physical results are independent on the lattice
orientation in the $L\rightarrow{}\infty$ limit). 

In the vicinity of zero energy ($|E|\rightarrow{}0$), there are four
solutions of the above
equation corresponding to four Dirac cones: the {\em central cone},
located at ${\bf p} = (p_x,p_y) = (0,0)$, and three {\em satellite cones},
located (in polar coordinates) at $p = { \gamma_1{}v_3}/{v_0^2}$,
$\varphi = {0,\ 2\pi/3,\ 4\pi/3}$. 
Below, we calculate the transmission of the system assuming that the states
corresponding to different Dirac cones do not interfere among themselves.
Physically, such a~supposition corresponds to the conditions for
the energy and system sizes
\begin{equation}
\label{condenwl}
  |E|,\ \frac{\hbar{}v_3}{L},\ \frac{\hbar{}v_3}{W} \ \ll{}\ E_L,
\end{equation}
where the Lifshitz energy $E_L=\frac{1}{4}\gamma_1(v_3/v_0)^2$.
For $\gamma_0=3.16\,$eV, $\gamma_1=0.381\,$eV, and $\gamma_3=0.3\,$eV
\cite{Kuz09}, we have $E_L\approx{}1\,$meV, and the last two conditions
in Eq.\ (\ref{condenwl}) are equivalent to
$L,\,W\gg{}4\l_\perp\gamma_3/\gamma_0=75\,$nm. 
%

Expanding the dispersion relation given by Eqs.\ (\ref{disp0a}) and
(\ref{disp0b}) up to the second order around ${\bf p} = (0,0)$, we obtain
\begin{equation}
\label{encencone}
  E^2 = v_3^2 \left( p_x^2 + p_y^2 \right).
\end{equation}
Thus, the central Dirac cone has isotropic dispersion relation, closely
resembling the dispersion relation following from the monolayer graphene
Hamiltonian [see Eq.\ (\ref{hammlg}) in the main text]; in fact, 
the only difference is the proportionality coefficient $v_3$ instead of $v_F$.

Now, we write down the effective {\em single-cone} Hamiltonian, 
corresponing to the dispersion relation given by Eq.\ (\ref{encencone})
\begin{equation}
  H_{\rm central} = \begin{pmatrix}
  0 & v_3\pi \\
  v_3\pi^\dagger & 0
  \end{pmatrix}.
\end{equation}
Solving the scattering problem for a rectangular sample described by the above
Hamiltonian with heavily (infinitely) doped leads one gets the formula for
transmission coefficient as a function of the transverse momentum
($k_y=p_y/\hbar$)
\begin{equation}
  T(k_y) = \frac{1}{\cosh^2(k_yL)}. 
\end{equation}
For the periodic boundary conditions, the transverse momentum gets quantized
values, $k_y^{(j)}=2\pi{}j/W$, with $j=0,\pm{}1,\pm{}2,\dots$.
For $W\gg{}L$, one can approximate the sum over $k_y^{(j)}$ by an integral, 
obtaining the total transmission
\begin{equation}
  \sum_jT(k_y^{(j)})\approx{} \frac{W}{2\pi}
  \int\limits_{-\infty}^{+\infty} T(k_y) {dk_y} =
  \frac{1}{\pi}\frac{W}{L}.
\end{equation}

Next, we expand (up to the second order) the dispersion relation around
${\bf p} = (\gamma_1 v_3/v_0^2, 0)$
(i.e., the satellite Dirac cone at $\varphi=0$), arriving to 
\begin{equation}
  E^2 =  \frac{ v_3^2 }{(1+(v_3/v_0)^2)^2}(p_x^2+9p_y^2).
\end{equation}
The corresponding single-cone Hamiltonian reads
\begin{equation}
  H_{\rm satellite}^{(\varphi=0)} = \begin{pmatrix}
  0 & v_3(p_x +i 3p_y) \\
  v_3(p_x -i 3p_y) & 0
  \end{pmatrix}.
\end{equation}
This time, solving the scattering problem for a rectangular sample we get
the transmission coefficient 
\begin{equation}
T(k_y) = \frac{1}{\cosh^2(3 k_yL)}, 
\end{equation}
and the integration over $k_y$ leads to
\begin{equation}
  \frac{W}{2\pi}
  \int\limits_{-\infty}^{+\infty} T(k_y) {dk} =
\frac{1}{3\pi}\frac{W}{L}. 
\end{equation}

The calculations for remaining Dirac cones at $\varphi\neq{}0$ are more
involving, yet straightforward. Generalizing the above reasoning for
${\bf p} =\gamma_1{}(v_3/v_0^2) \left(\cos\varphi, \sin\varphi\right)$, we get
\begin{multline}
  E^2 =  \frac{ v_3^2 }{(1+(v_3/v_0)^2)^2}
  \Big[\,p_x^2+p_y^2 \\
  + 8\left(p_x \sin\varphi + p_y\cos\varphi\right)^2\Big],
\end{multline}
and 
\begin{equation}
H_{\rm satellite}^{(\varphi)} = \left(\alpha \sigma_x + \beta \sigma_y\right),
\end{equation}
where
\begin{align}
  \alpha &= p_x\cos\varphi - p_y\sin\varphi, \\
  \beta &= -3p_x\sin\varphi - 3p_y\cos\varphi. 
\end{align}
Finally, we have
\begin{equation}
T(k_y) = \frac{1}{\cosh^2\left[3 k_yL/\left(5-4 \cos(2\varphi)\right)\right]},
\end{equation}
and
\begin{equation}
\frac{W}{2\pi}\int\limits_{-\infty}^{+\infty} T(k_y){dk} =
\frac{5-4\cos(2\varphi)}{3\pi}\,\frac{W}{L}.
\end{equation}

Summing up the contributions from all four Dirac cones, we obtain total
transmission
\begin{equation}
T_{\rm total} = \frac{6\,W}{\pi{}L}.
\end{equation} 
Substituting the above into Eq.\ (\ref{sigzerte}) in the main text,
we obtain $\sigma=3\sigma_0$ in the physical units.
Remarkably, the result is independent on the lattice orientation, as we have
$\cos(\varphi) + \cos(\varphi+4\pi/3)+\cos(\varphi-4\pi/3) =0$ for any real
value of $\varphi$.
(Notite that the summation of independent contributions from four Dirac cones,
performed above, instantly reproduces the limit of $L,W\rightarrow\infty$.)

Similarly, for the Fano factor we have
\begin{equation}
  F = 1-\frac{\displaystyle\sum_{\rm cones}\int{}dk_y\left[\,T(k_y)\,\right]^2}{%
  \displaystyle\sum_{\rm cones}\int{}dk_y\,T(k_y)}
  =\frac{1}{3}.
\end{equation}

$\ $

\section{
  The effect of {\boldmath$\gamma_4\neq{}0$} tunneling in the absence of
  trigonal warping ({\boldmath$\gamma_3=0$}) 
}

In this section, we consider the case complementary to the analyzed
in Appendix~A. 

The Hamiltonian given by Eq.\ (\ref{hamblg}) in the main text, 
for $\delta_{AB}=\gamma_3=0$, reduces to
\begin{equation}
  \label{hamblg4}
  H_{\rm BLG} =
  \left(
  \begin{array}{cccc}
    0 & v_0\pi & \gamma_1 & -v_4\pi^\dagger \\
    v_0\pi^\dagger & 0 & -v_4\pi^\dagger & 0 \\
    \gamma_1 & -v_4\pi & 0 & v_0\pi^\dagger \\
    -v_4\pi & 0 & v_0\pi & 0  \\
  \end{array}
  \right), 
\end{equation}
leading to the two low-energy bands (with $E=0$ for $p=0$) and the two
high-energy bands (with $E=\pm{}\gamma_1$ for $p=0$) in the dispersion
relation.
For low energies, one can write down the effective two-band Hamiltonian
(see Ref.\ \cite{Mac13})
\begin{equation}
H_{\rm 2band} = \frac{1}{2m}\left(\begin{array}{cc}
  \mu_4 \pi\pi^\dagger &  -\left(\pi\right)^2 \\
  -\left(\pi^\dagger\right)^2
    &  \mu_4 \pi^\dagger\pi
\end{array}\right), 
\end{equation}
where $m=\gamma_1/2v_0^2$ and $\mu_4=4mv_0v_4/\gamma_1=2v_4/v_0$. 

Now, we follow the approach proposed by Katsnelson in Ref.\ \cite{Kat06b},
performing the mode matching for two interfaces between heavily- and
weakly-doped areas (the leads and the sample), separated by a~distance $L$. 
For a~fixed but finite doping in the leads (quantified by the Fermi wavenumber
$k_F$), elementary analysis leads to the following formula for transmission
coefficient for a~given transverse wavenumber $k_y$ (conserved at both
the interfaces)

\begin{widetext}
\begin{equation}
\label{tky2band}
  T(k_y) = 
  \frac{16 \zeta ^2 \Big\{(-1-\mu_4) {k_F L} \,\zeta  \cosh(\zeta )
  -\mu_4 \left[(-1-\mu_4){k_F L} + 2 \zeta ^2\right] \sinh(\zeta )\Big\}^2}%
  {4 \zeta^2 \left[\zeta \cosh(2 \zeta ) -
  \mu_4 {k_F L} \sinh(2 \zeta )\right]^2+
  \left({k_F L}\right)^2 \Big\{-2{k_F L}\,\zeta^2 +
  \mu_4^2 {k_F L} \left[\cosh(2 \zeta )-1\right]
  -2 \mu_4 \zeta  \sinh(2 \zeta )\Big\}^2},
\end{equation}
where we have defined $\zeta=k_yL$. 
\end{widetext}

Changing the variables according to $T(k_y)\equiv{}T(\zeta,L)$ we find
that the Landauer-B\"{u}ttiker conductivity, for a~fixed $W/L\gg{}1$,
is bounded by
\begin{equation}
\label{sigboundmu4}
  \sigma(L)=\frac{1}{2\pi}\int{}d\zeta{}\,T(\zeta,L)\lesssim{}
  \frac{\text{const}}{L^2}\ \ \ \ (\text{for }\ \mu_4\neq{}0), 
\end{equation}
vanishing in the $L\rightarrow{}\infty$ limit.
For $\mu_4=0$, the conductivity $\sigma(L)\approx{}(\pi/4)\sigma_0$, 
and the Fano factor $F\approx{}1-2/\pi$ for $L\gg{}l_\perp$,
being numerically close to the results by Snyman and Beenakker, see
Ref.\ \cite{Sny07}. 

\begin{figure}
  \includegraphics[width=0.8\linewidth]{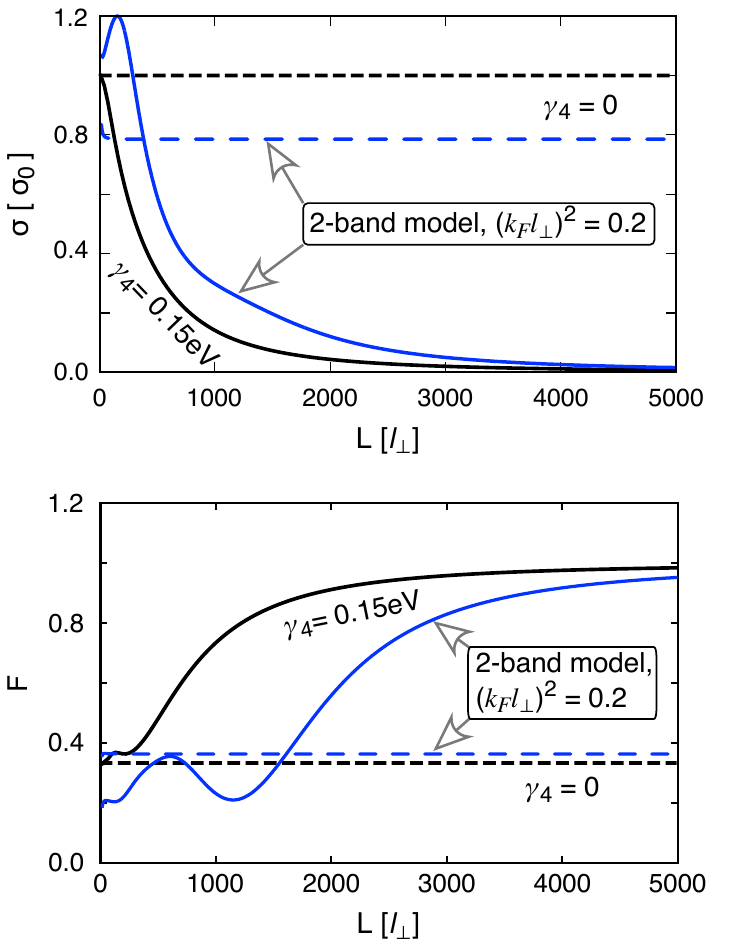}
\caption{ \label{sigfan2band}
  Conductivity (top) and the Fano factor (bottom) obtained from Eq.\
  (\ref{tky2band}) for $(k_Fl_\perp)^2=0.2$ and $\gamma_4=0.15\,$eV
  (blue solid lines), or $\gamma_4=0$ (blue dashed lines).
  The corresponding results for the four-band model are reproduced from
  Figs.\ \ref{conductivity}(c) and \ref{fanofactor}(c) for a~comparison
  (black solid and black dashed lines). 
}
\end{figure}

The approximate upper bound for  $\mu_4\neq{}0$, given in Eq.\
(\ref{sigboundmu4}),
is further supported with the numerical results presented in
Fig.\ \ref{sigfan2band}, where we have set the doping in the leads
such that $(k_F{}l_\perp)^2=0.2$ (after Ref.\ \cite{Sny07}), and $W/L=20$.
Numerical calculations for the full four-band model given
the Hamiltonian $H_{\rm BLG}$ (\ref{hamblg4})
leads to a~noticeably faster, but also a~power-law decay of the conductivity, 
which can be approximated as $\sigma(L)\propto{}L^{-2.0}$ for
$L\gtrsim{}1000{}\,l_\perp$.

\begin{thebibliography}{}
\bibitem{anglefoo}
The limit of $\sigma(L)\rightarrow{}3\sigma_0$ for $L\rightarrow{}\infty$
appears for a~generic crystallographic orientation of a~sample.
In a~particular case when the current is passed precisely along a~zigzag
direction, the limit of $(7/3)\,\sigma_0$ is predicted; 
see Ref.\ \cite{Mog09}.
\bibitem{gam2foo}
The next-nearest layer parameter $\gamma_2$ is not present in bilayer graphene,
see Refs.\ \cite{Mcc57,Slo58}.
\bibitem{nonrelfoo}
For Schr\"{o}dinger electrons, incident on a~rectangular barrier at zero
energy (with respect to the top of the barrier), the transmission probability
reads 
$T_n=\left[\cosh^2Lq_n+\left(k_\infty^2/2q_n^2-1\right)\left(k_\infty^2/q_n^2-1\right)^{-1}\sinh^2Lq_n\right]^{-1}$, with $q_n=2\pi{}n/W$ ($n=0,1,2,\dots$) being the quantized transverse wavebumber for the periodic boundary conditions, and $k_\infty$ denoting the wavenumber in heavily-doped leads. For $k_\infty\rightarrow\infty$, we have $T_n\propto{}(q_n/k_\infty)^2$, vanishing for any $n$.
\end{thebibliography}
\end{document}